\documentstyle[aps,prc]{revtex}
\input{psfig}
\begin{document}
\draft
\title{Shell Model Study of the Neutron-Rich Nuclei around N=28}
\author{{\bf J. Retamosa$^{*}$, E. Caurier$^{*}$, F. Nowacki$^{**}$
        and A. Poves$^{***}$}\\
      CRN IN2P3-CNRS/Universit\'e Louis Pasteur BP20\\ 
      F-67037 Strasbourg-Cedex, France.\\ 
      $^{**}$ GANIL IN2P3-CNRS BP5027\\
      F-14021 Caen-Cedex, France\\  
      $^{***}$ Departamento de F\'{\i}sica Te\'orica C-XI\\
      Universidad Aut\'onoma de Madrid\\ 
      E-28049 Madrid, Spain.} 
\date{\today} 
\begin{center} 
\maketitle 
\begin{abstract} 
  We describe the properties of the neutron rich nuclei around N=28
  in the shell mode framework. The valence space comprises the $sd$
  shell for protons an the $pf$ shell for neutrons without any
  restriction. Good agreement is found with the available experimental
  data. The N=28 shell closure, even if  eroded due to the large neutron
  excess, persists. The calculations predict that  
  $^{40}$S and  $^{42}$S are deformed with $\beta=0.29 $ and
  $\beta=0.32$  respectively.
  
\end{abstract} 
\end{center} 
 
\section{Introduction} 
 
The exploration of the behaviour of the nucleus under extreme
conditions;  high spin, vicinity to the drip lines, finite temperature,
etc. is a major source of new insights in nuclear structure.  One of
the most important questions raised by the study of the drip lines is
whether the basic shell order evolves with the neutron (proton) excess
An example of this problem is given by the N=20 very neutron rich
nuclei where the strong anomalies
detected\cite{Th:75,Det:78,Hub:78,Gui:84} were explained as a collapse
of the standard N=20 shell closure. Extended shell model calculations
\cite{Po:94,War:90,Fuk:92} showed that the inversion of the spherical
closed-shell configurations and the 2p-2h collective intruders could
account for the large binding energies and half-lives, unexpected
ground-state spins, and high level densities at low excitation
energies.  Also, mean-field calculations using the Skyrme interaction
predicted a sharp transition from spherical to prolate
shapes\cite{Cam:75}

Recently, there has been an increase of  interest in the N=28 isotones
far from the stability, motivated by;  i) the possible existence of
anomalies in the shell closures, as already found
in N=20 and ii) the unusual $^{48}$Ca/$^{46}$Ca abundance ratio 
measured in the solar system or the Ca-Ti-Cr anomalies observed in some
meteoritic inclusions.  

Sorlin et al. \cite{Sor:94} undertook the study of the $\beta$-decay
and $\beta$-delayed neutron emission probabilities for $^{44}$S and  
$^{45-47}$Cl. Their measured half-lives were much larger than those
predicted by TDA\cite{Kl:84} or QRPA\cite{St:90} calculations.  These  
discrepancies were attributed to unexpected shape transitions in  
the region. Mass formulas such as the finite-range liquid drop\cite{Mo:88}, 
the finite-range droplet\cite{Mo:92} or the Extended Thomas-Fermi with 
shell corrections \cite{Pea:91} predict the existence of shape       
coexistence around $^{44}$S. In all the cases the energy surfaces are
very soft with close lying minima corresponding to different deformations.

Similar conclusions have been drawn by Werner et al\cite{Wer:94}. 
Relativistic mean field calculations and non-relativistic ones using
the Skyrme forces produce 
very flat energy surfaces for the Sulphur isotopes, with several 
minima separated by energy barriers of just a few hundred keV. 
In this his situation it seems necessary to go beyond
the mean-field using, for instance, the generator coordinate method. 
 
In this paper we discuss all the neutron rich nuclei with N$\ge 20$ 
and $14\le$Z$\le 20$ in the Shell Model framework. This is, to some 
extent, complementary to the other methods mentioned above and can 
give new  information  on the structure of these nuclei. The paper is 
organized as follows. Section II gives a brief description of the 
model, discussing the choice of the valence space, the effective 
interaction  and the coulomb corrections. Section III compares our 
results with the experimental data.  In section IV we examine the 
behaviour of the N=28 shell closure far from the stability. 
Section V deals with deformation in heavy Sulphur isotopes.

\section{The Model} 
 
\subsection{The Valence Space} 
 
It consists of the full $sd$ shell for Z-8 protons and the full $pf$
shell for N-20 neutrons. Although the calculations reach very large
dimensionalities, they can be performed without truncations in all
cases.  We do not include the intruder states that could be formed by
exciting particles from the $sd$ to the $pf$ shell. These kind of
states, that appear at low excitation energy around $^{40}$Ca, should
be less important near N=28.  The reason is twofold; $sd$ protons get
more and more bound with increasing neutron excess then unfavouring
proton intruders; and the correlation energy of the configurations
with neutron intruders do not gain much energy compared to the
standard ones (the opposite is true in e.g. $^{32}$Mg).  This is in
agreement with the observed spectra. The existence of many collective
low lying states near N=Z=20 contrast with their absence for
N$\simeq28$, Z$>20$.

\subsection{The interaction}  
 
The construction of the effective hamiltonian is critical in any shell
model description. It is well known that the saturation properties of
the realistic interactions must be corrected phenomelogically in order
to have a good description of the nuclear structure\cite{Abz:91}.
Unfortunately, the experimental data for the very neutron rich nuclei
with $14\le $Z$\le 18$ are scarce. Even if the masses are known over a
relatively wide region, at N=28 they have only been measured for
nuclei with Z$\ge19$. As an example the masses of the Sulfur isotopes
are only known up to N=26.  Half-lives are only known for $^{44}$S and
$^{45-47}$Cl.  The level schemes of Ar and Cl with $20\le N \le 23$
are partly established.  The situation is clearly better for the K
isotopes where the evolution of of the $3/2^{+},1/2^{+}$ doublet with
the neutron number is known up to N=28\cite{Wal:90}.

Our starting interaction has three parts. The USD interaction of
Wildenthal\cite{Wil:83} is used for the particles lying in the
sd-shell. A modified version of the Kuo-Brown interaction, denoted
KB$^{'}$ in ref \cite{Pov:81}, gives the two body matrix elements for
the $pf$ shell particles. This interaction is better suited to this
valence space that the KB3 interaction that has been used in
conventional $pf$ shell calculations \cite{Cau:94}.  Furthermore, when
used in the lower part of the $pf$ shell it also gives good
spectroscopic results.  Finally, the cross-shell interaction is the
G-matrix of Lee, Kahanna and Scott (LKS)\cite{LKS}.  This is the part
of the interaction whose monopoles we have proceeded to modify
phenomenologically.  The data used in the fit are:

\begin{itemize}
\item The natural parity states of nuclei with few $pf$ particles and sd
      holes.

\item The evolution of the $3/2^{+},1/2^{+}$ doublet along the K isotopes.

\item The position of the 5/2$^{+}$ states in $^{47}$K.

\end{itemize}

The experimental data cited above are not enough to fix all the
monopole parameters. They essentially fix the interaction among
neutrons in the 1f$_{7/2}$ and 2p$_{3/2}$ orbits with the $sd$
protons.  On the contrary, the interaction with 1f$_{5/2}$ and
2p$_{1/2}$ particles is not well determined. In our choice of the
final interaction $^{41}$Ca and $^{35}$Si have very similar single
neutron spectra as predicted in Duflo and Zuker \cite{Du:95} mass
formula.  There exists other possible choices of this part of the
interaction that induce important modifications in the structure of
heavy Silicons and to a minor extent in that of the heavy Sulphurs.
 
The 2-body matrix elements incorporate a tiny mass dependence and are
written as:

\begin{equation} 
  V = \frac{\omega(A)}{\omega(A_0)} V_0 = \frac{<r^2(A_0)>}{<r^2(A)>} V_0
\end{equation}
 
where $A_0=40$ and $\hbar\omega_0$=11 MeV are used to calculate the
LKS G-matrix.  Instead of the usual law $<r^2> = r_0 A^{1/3}$ fm, we
borrow from reference \cite{Du:95} a modified expression, which is
better adapted to large mass regions.

\begin{equation} 
  <r^2> = r_0
  \left[A\left(1-\zeta\left(\frac{N-Z}{A}\right)^{1/2}\right) \right
    ]^{1/3}.
\end{equation} 

leading to the following expression for the 2-body matrix elements

\begin{equation} 
  V = \left[\left(\frac{A_0}{A}\right) \left(
  1-\zeta\left(\frac{N-Z}{A} \right)^{1/2}\right)\right]^{1/3}V_0
\end{equation} 

with $\zeta=0.42$ and $A_0=40$.

\subsection{Coulomb Energies}

The absolute energies predicted with the precedent interaction are
relative to an $^{16}O$ core and do not include the coulomb energy of
the protons.  Thus, to compare our predictions with the experiment,
they must be corrected with the following prescription:
 
\begin{equation} 
  E_{i}(N,Z) = E_0+E_c(N,Z)+E_{i}^{SM}(N,Z)
\end{equation}

where $E_0$ is the inert core energy (127.62 MeV), $E_c(N,Z)$ is the
coulomb energy relative to the core and $E^{SM}_{i}$ is the nuclear
energy obtained with our interaction.  The $E_c$ energy, which is
assumed to depend only on N and Z, has been calculated with the Antony
and Pape formula\cite{AP:88}

\begin{equation}
  E_c(A,Z) = \begin{array}{ll}
  E_c(A,Z-1)+1.44(Z-1/2)/A^{1/3}-1.02&Z>Z_s\\ 
  E_c(A,Z+1)-1.44(Z+1/2)/A^{1/3}+1.02&Z<Z_s \end{array}
\end{equation}

where $Z_s = A/2$ or $(A+1)/2$ and the coulomb correction $E_c(Z_s)$
along the stability line is given by

\begin{equation} 
  E_c(Z_s) = \begin{array}{ll} 0.162Z^2+0.95Z-18.25&Z\le 20\\ 
  0.125Z^2+2.35Z-31.53&Z> 20 \end{array}
\end{equation}

which agrees with the Chung-Wildenthal coulomb correction\cite{War:90}
when $Z_s\le20$ and with that of reference\cite{Cau:94} for $Z_s >
20$.

\section{Results and Discussion}

\subsection{Comparison with the data.}

In this section we compare compare the theoretical predictions with
the experimental data.  The shell model matrices are dealt with by the
code {\sl ANTOINE} \cite{Cau:90} a very fast and efficient
implementation of the Lanczos algorithm . The USD predictions for the
$sd$-shell nuclei have been largely reported \cite{Wil:83}. However,
we shall include $sd$ nuclei in the $S_{2N}$ plots to stress the
continuity of the shell model description across the N=20 magic
number.

Table \ref{t1}, contains the predicted and experimental energies of
the $2^{+}$ states, that show an excellent agreement.  The
interaction, adjusted to reproduce the behaviour of the 3/2$^{+}$,
1/2$^{+}$ levels along the odd-A K, predicts for the doublet a similar
behaviour in the neutron rich Chlorine and Phosphorous (see table
\ref{t2}).  The crossing of these two levels is due to the gradual
degeneracy of the 2s$_{1/2}$ and 1d$_{3/2}$ orbitals as the occupation
of the 1f$_{7/2}$ shell increases. In the P isotopes, the 1$d_{5/2}$
orbit is nearly closed, consequently the evolution of the doublet
gives us an idea of the relative position of the two orbitals.  As we
shall comment later, their relative possition is determinant for the
development of deformation in the heavy Sulphurs.

\begin{figure*}[top]
\begin{center}
\caption[]{Theoretical and experimental natural parity states for $^{40}K$}
\label{f1}
\vspace*{0.5cm}
\leavevmode
\psfig{file=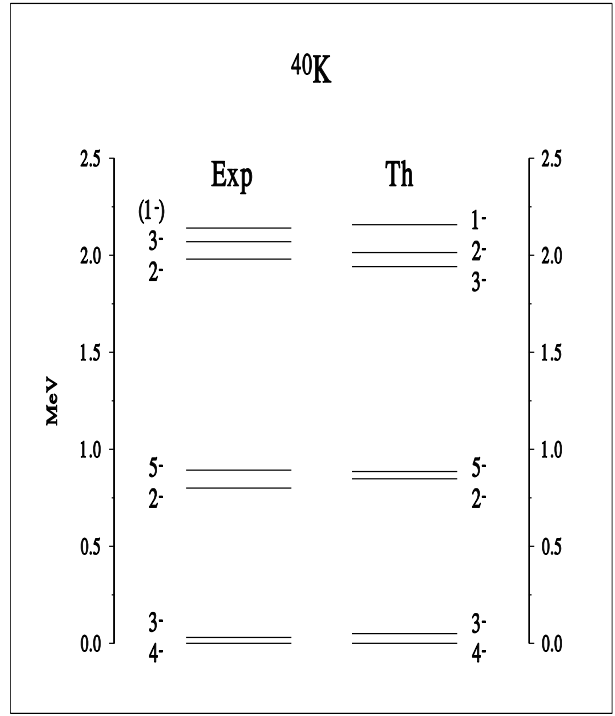,height=9.5cm ,width=8cm}

\end{center}
\end{figure*}

\begin{figure*}[bot]
\begin{center} 
\caption[]{Same as figure \ref{f1} for $^{41}K$}
\label{f2} 
\vspace*{0.5cm} 
\leavevmode 
\psfig{file=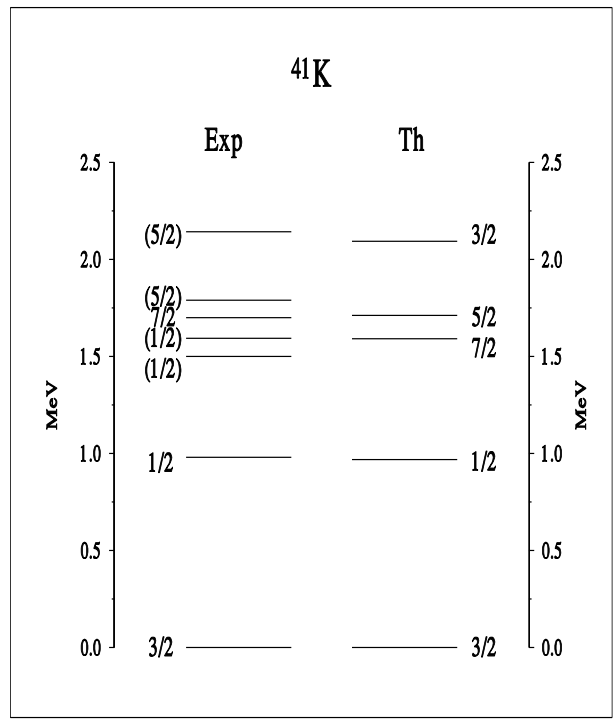,height=9.5cm ,width=8cm}
\end{center} 
\end{figure*}

\begin{figure*}[top]
\begin{center} 
\caption[]{Same as figure \ref{f1} for $^{42}K$}
\label{f3} 
\vspace*{0.5cm} 
\leavevmode 
\psfig{file=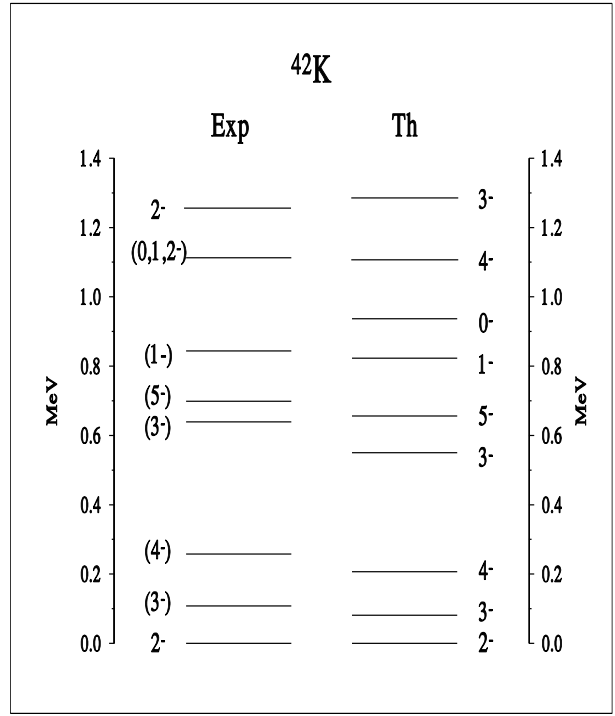,height=9.5cm ,width=8cm}
\end{center} 
\end{figure*}

\begin{figure*}[bot]
\begin{center} 
\caption[]{Same as figure \ref{f1} for $^{39}ar$}
\label{f4}
\vspace*{0.5cm}
\leavevmode
\psfig{file=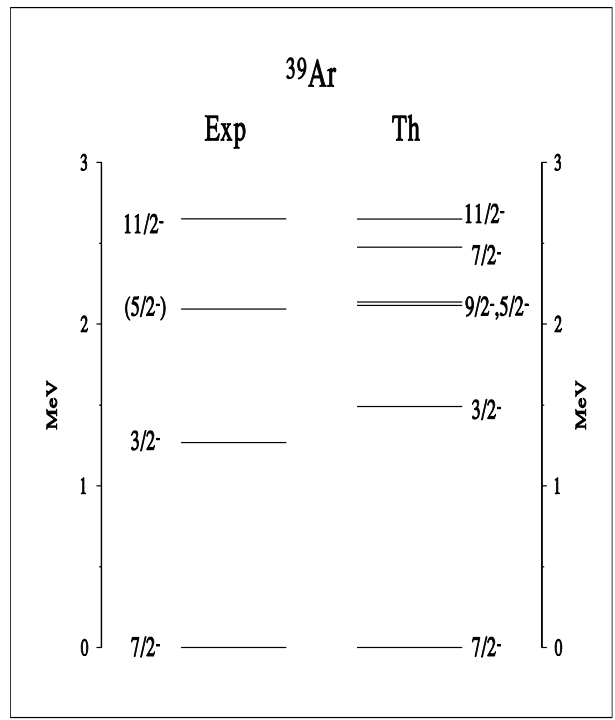,height=9.5cm ,width=8cm}
\end{center} 
\end{figure*}

\begin{figure*}[top]
\begin{center}
\caption[]{Same as figure \ref{f1} for $^{40}Ar$}
\label{f5}
\vspace*{0.5cm}
\leavevmode
\psfig{file=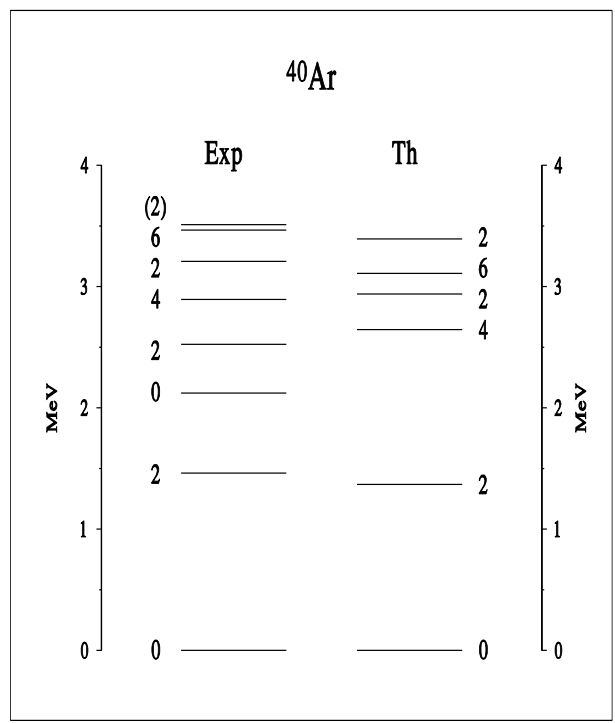,height=9.5cm ,width=8cm}
\end{center}
\end{figure*}

\begin{figure*}[bot]
\begin{center}
\caption[]{Same as figure \ref{f1} for $^{41}Ar$}
\label{f6}
\vspace*{0.5cm}
\leavevmode
\psfig{file=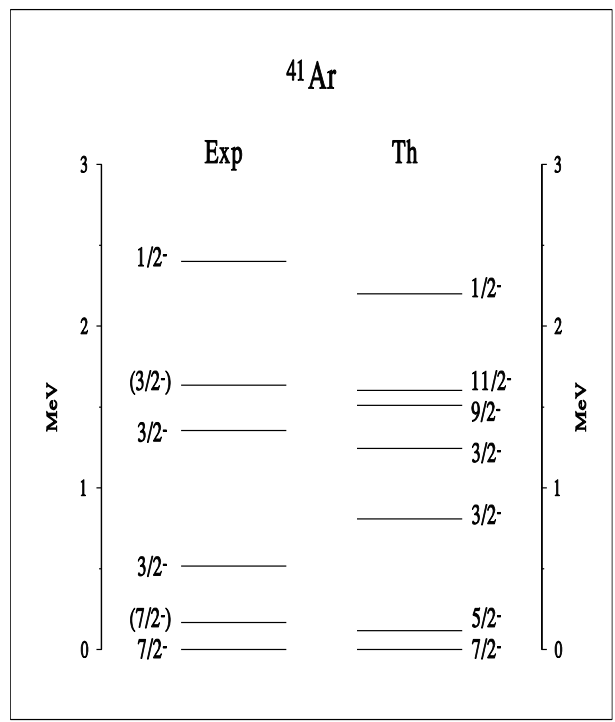,height=9.5cm ,width=8cm}
\end{center}
\end{figure*}

\begin{figure*}[top]
\begin{center}
\caption[]{Same as figure \ref{f1} for $^{38}Cl$}
\label{f7} 
\vspace*{0.5cm}
\leavevmode
\psfig{file=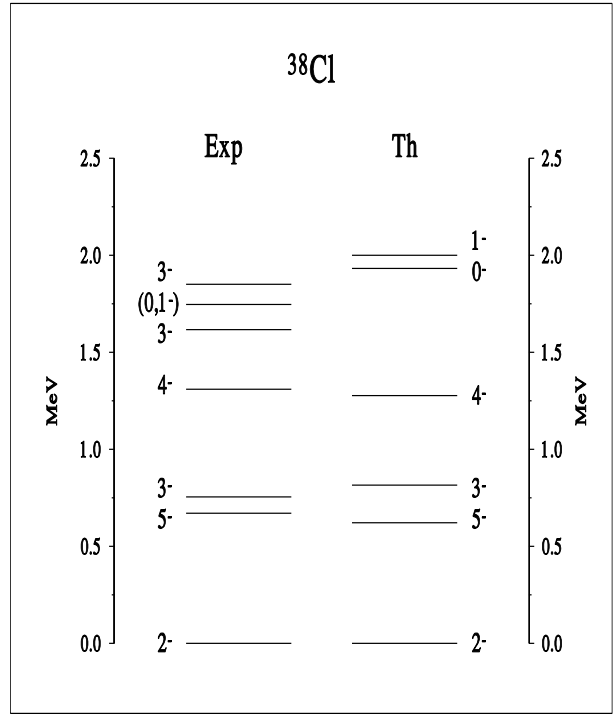,height=9.5cm ,width=8cm}
\end{center}
\end{figure*}

The spectroscopic information on the nuclei with Z$<20$ N$>20$ is
limited to a some K, Ar and Cl isotopes.  Figures \ref{f1} to \ref{f7}
show the theoretical and experimental spectra for the cases
--$^{40-42}$K, $^{39-41}$Ar and $^{38}$Cl-- for which the energy
levels are known. Yet, there are still many uncertainties in the
measured level schemes, specially the spin assignments. The
theory-experiment agreement is perfect for $^{40-41}$K and $^{40}$Ar.
The few states without theoretical counterpart are most probably
sd-intruders.  For the other nuclei the concordance is also quite
good.

In addition to these spectra, the half-lives of several Ar and S
isotopes are known. Their decays are dominated by allowed Gamow Teller
transitions.  The ground states of the father nuclei were calculated
in our $0 \hbar \omega$ space and the Gamow-Teller sum rule state was
determined in a larger space that includes all the $1p-1h$ in order to
recover the $3(N-Z)$ sum rule.  The strength functions were obtained
by the Whitehead method \cite{Whi:80}.  The phase space factors of
Wilkinson\cite{Wilk:74} were used to compute the half-lifes. Total
strengths and half-lives are compiled in table \ref{t3}. The agreement
is very satisfactory. The calculated half-life of $^{44}$S is very
sensitive to the $Q_{\beta}$.  We have used the theoretical value; if
we had employed the extrapolated number given in the Audi and Wapstra
compilation \cite{Audi}, the half-life would have increased to
150$ms$.

\begin{figure*}[bot]
\begin{center}
\caption[]{S$_{2n}$ energies for the Ca isotopes.}
\label{f8}
\vspace*{0.5cm}
\leavevmode
\psfig{file=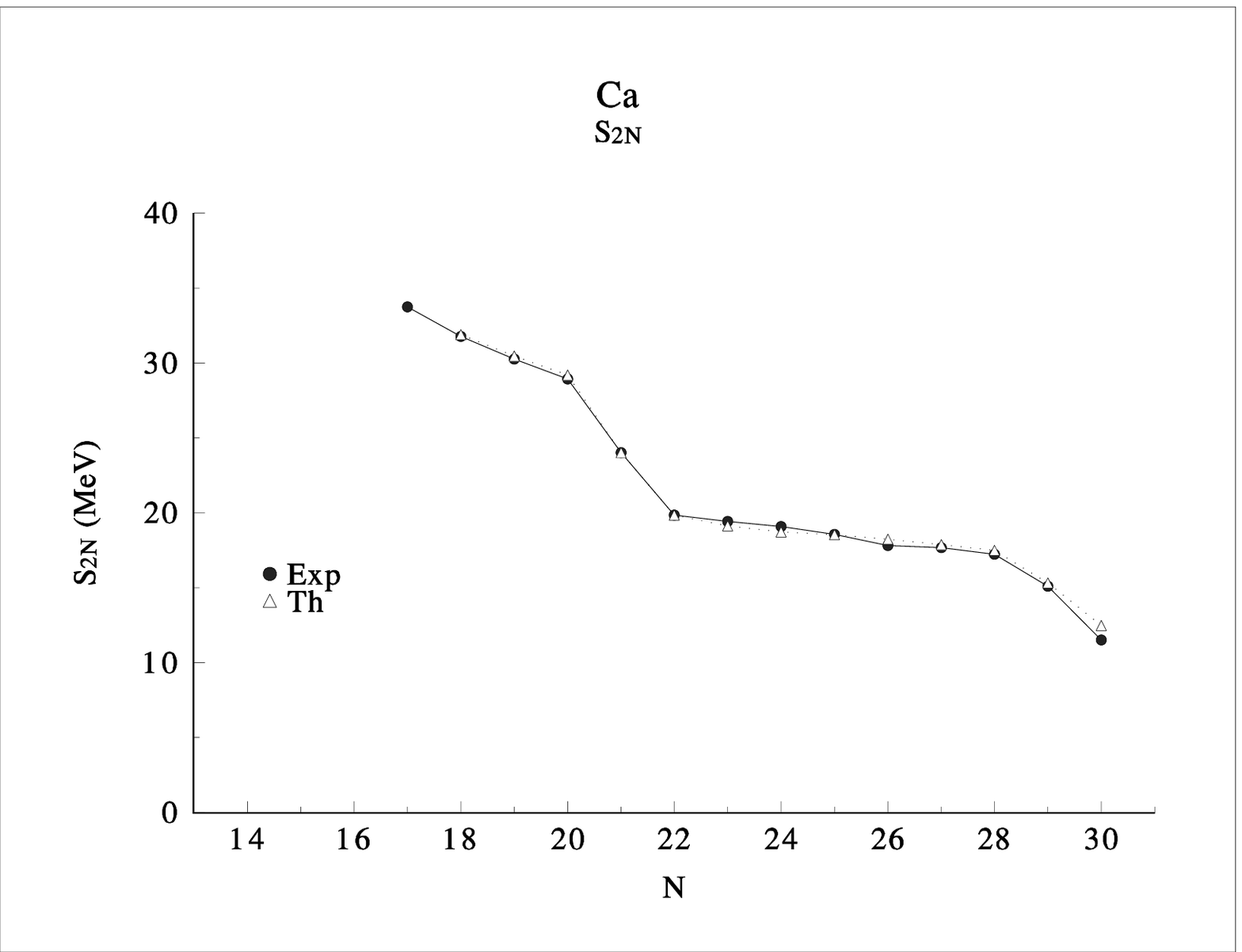,height=9.5cm ,width=8cm}
\end{center}
\end{figure*}

\begin{figure*}[top]
\begin{center}
\caption[]{S$_{2n}$ energies for the K isotopes.}
\label{f9}
\vspace*{0.5cm}
\leavevmode
\psfig{file=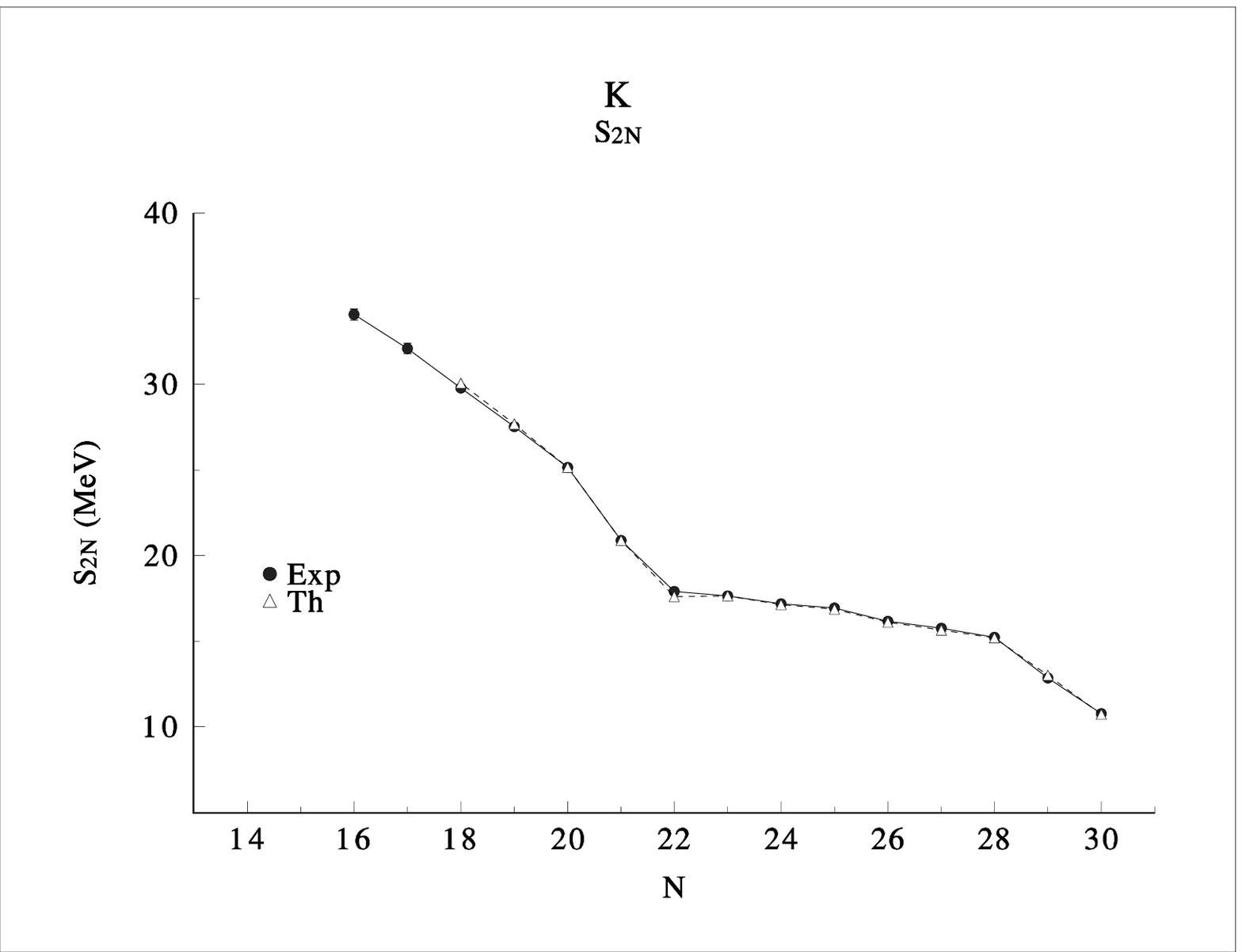,height=9.5cm ,width=8cm}
\end{center}
\end{figure*}

\begin{figure*}[bot]
\begin{center}
\caption[]{S$_{2n}$ energies for the Ar isotopes. The experimental
  points at the vertical line beyond are not
  measured quantities but numbers from Audi and Wapstra's extrapolations}
\label{f10}
\vspace*{0.5cm}
\leavevmode
\psfig{file=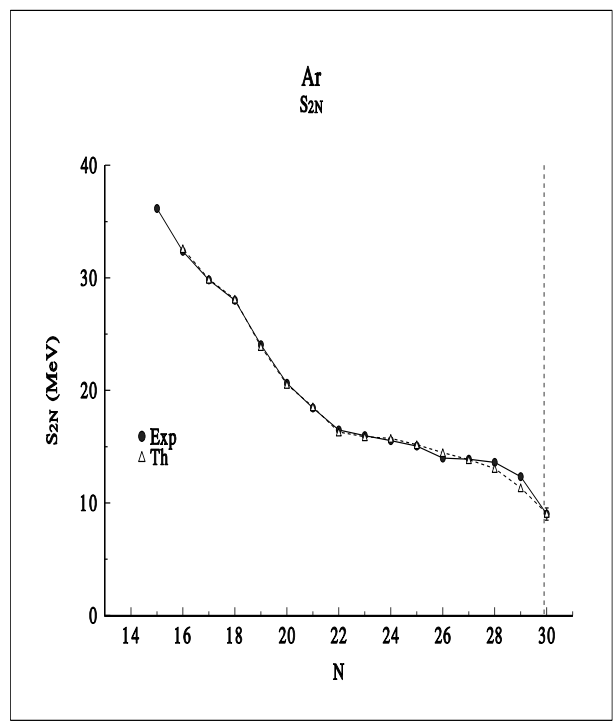,height=9.5cm ,width=8cm}
\end{center}
\end{figure*}

\begin{figure*}[top]
\begin{center}
\caption[]{S$_{2n}$ energies for the Cl isotopes. See caption to figure
           \ref{f10}. }
\label{f11}  
\vspace*{0.5cm} 
\leavevmode 
\psfig{file=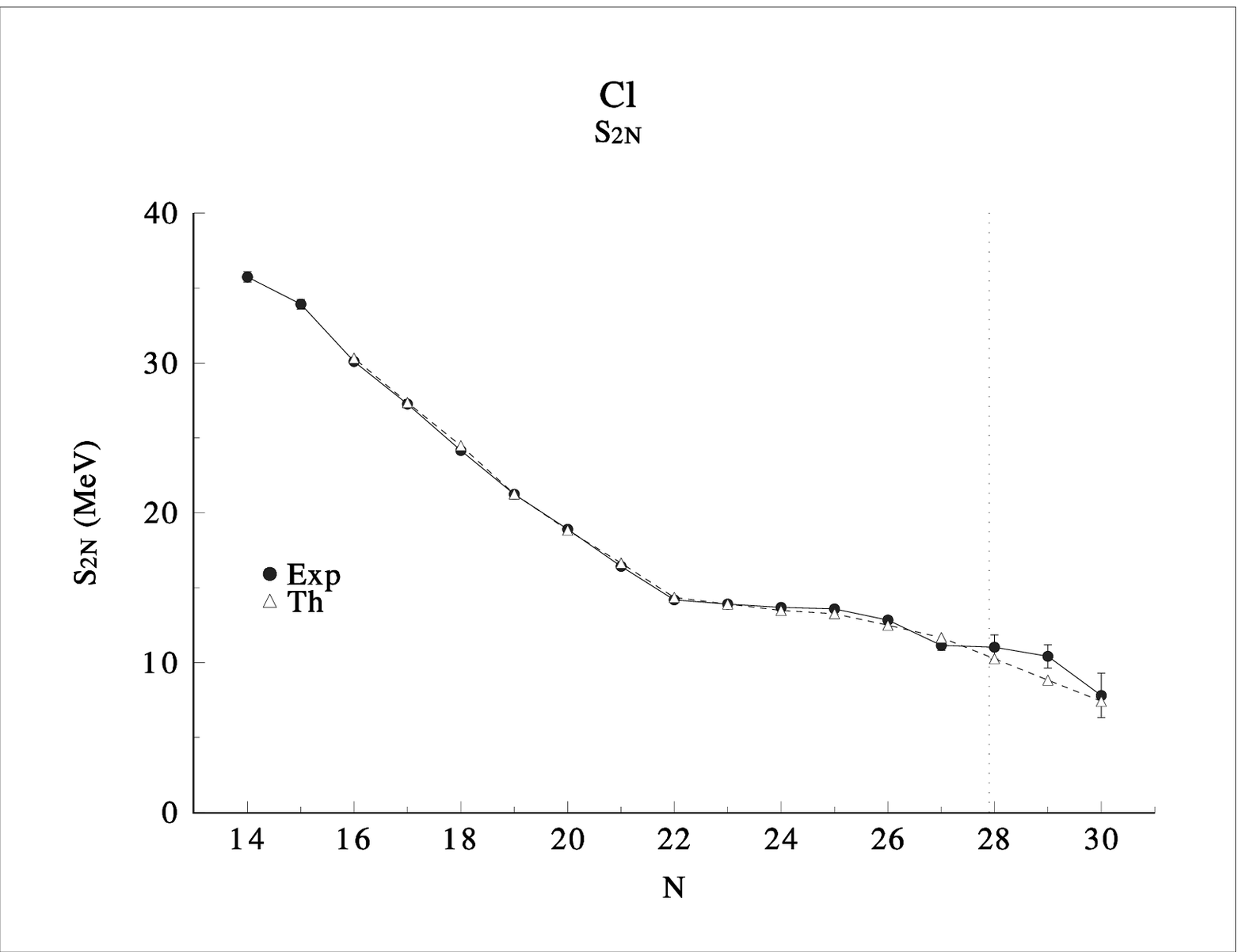,height=9.5cm ,width=8cm}

\end{center} 
\end{figure*}

\begin{figure*}[bot] 
\begin{center} 
\caption[]{S$_{2n}$ energies for the S isotopes. See caption to figure
           \ref{f10}.}
\label{f12}
\vspace*{0.5cm}
\leavevmode
\psfig{file=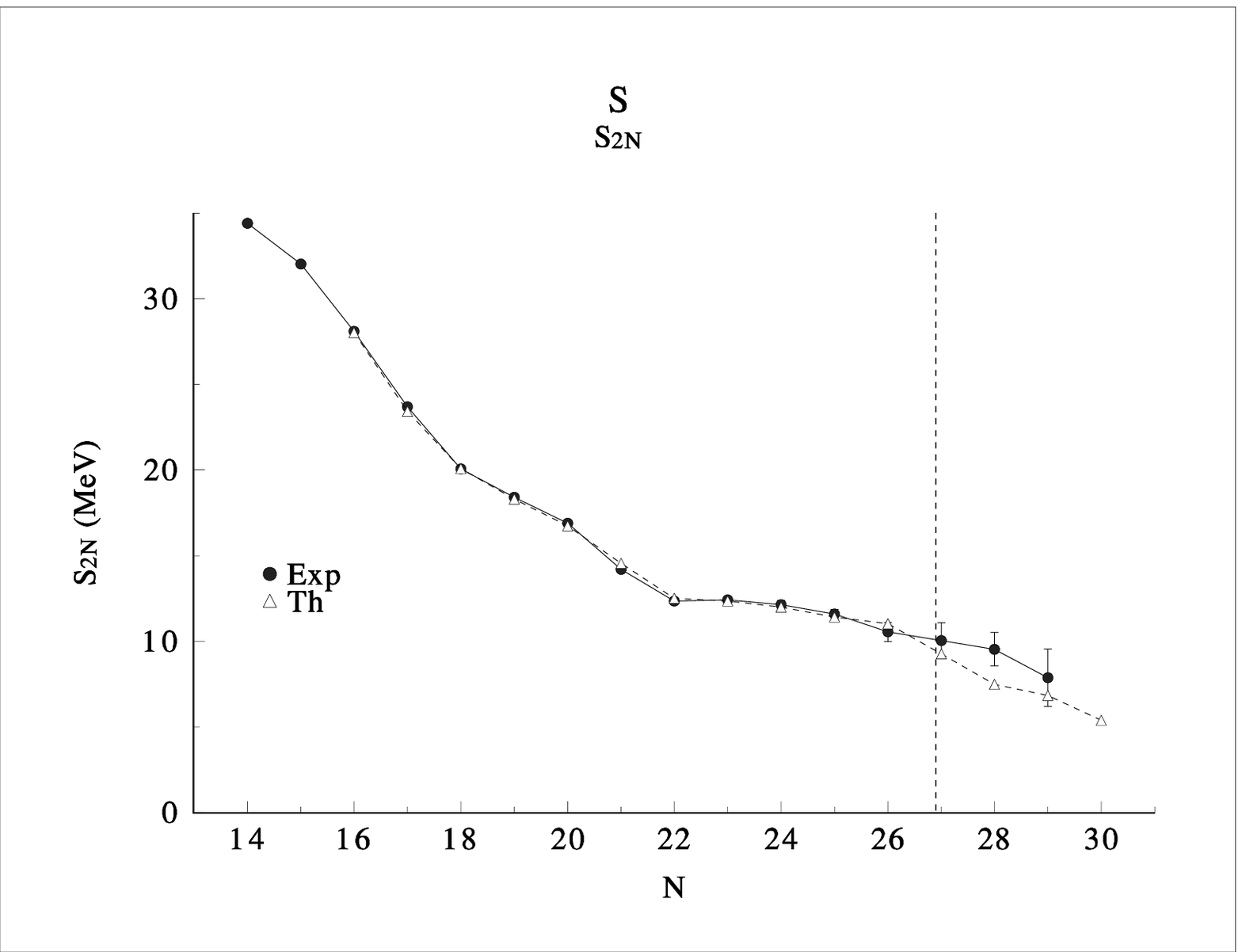,height=9.5cm ,width=8cm}
\end{center} 
\end{figure*}

The two neutron separation energies S$_{2N}$ are known over a large
region, unfortunately, for most Z's they do not reach N=28. The
agreement with the experimental results is good as it can be seen in
figures \ref{f8} to \ref{f12}. However, the discrepancies with the
extrapolated masses \cite{Audi} are important.

\subsection{Does the N=28 shell gap persist?}

To answer to this question we analyse our  predictions for the
S$_{2N}$, the excitation energies of the first $2^{+}$ states, and the
occupation numbers. Structural changes, such as spherical to deformed
shape transitions, should show up in the behaviour of these quantities
as a function of the neutron number.

The S$_{2N}$'s around N=28 have some common features for different
Z-values.  In three cases: Ca, K and Si, a sharp decrease of
S$_{2N}$ is observed at N=28. For the other isotopes, there are not
sharp slope changes, but  S$_{2N}$ still decreases. These results are
consistent with a  shell closure. This conclusion is opposite
to that coming from the mass systematics, that predicts
almost constant S$_{2N}$ around N=28, and thar could be
interpreted  as a signature of a new region of deformation.

\begin{figure*}[top]
\begin{center}
\caption[]{S$_{2n}$ energies for the P isotopes. See caption to figure
           \ref{f10}.}
\label{f13}
\vspace*{0.5cm}
\leavevmode
\psfig{file=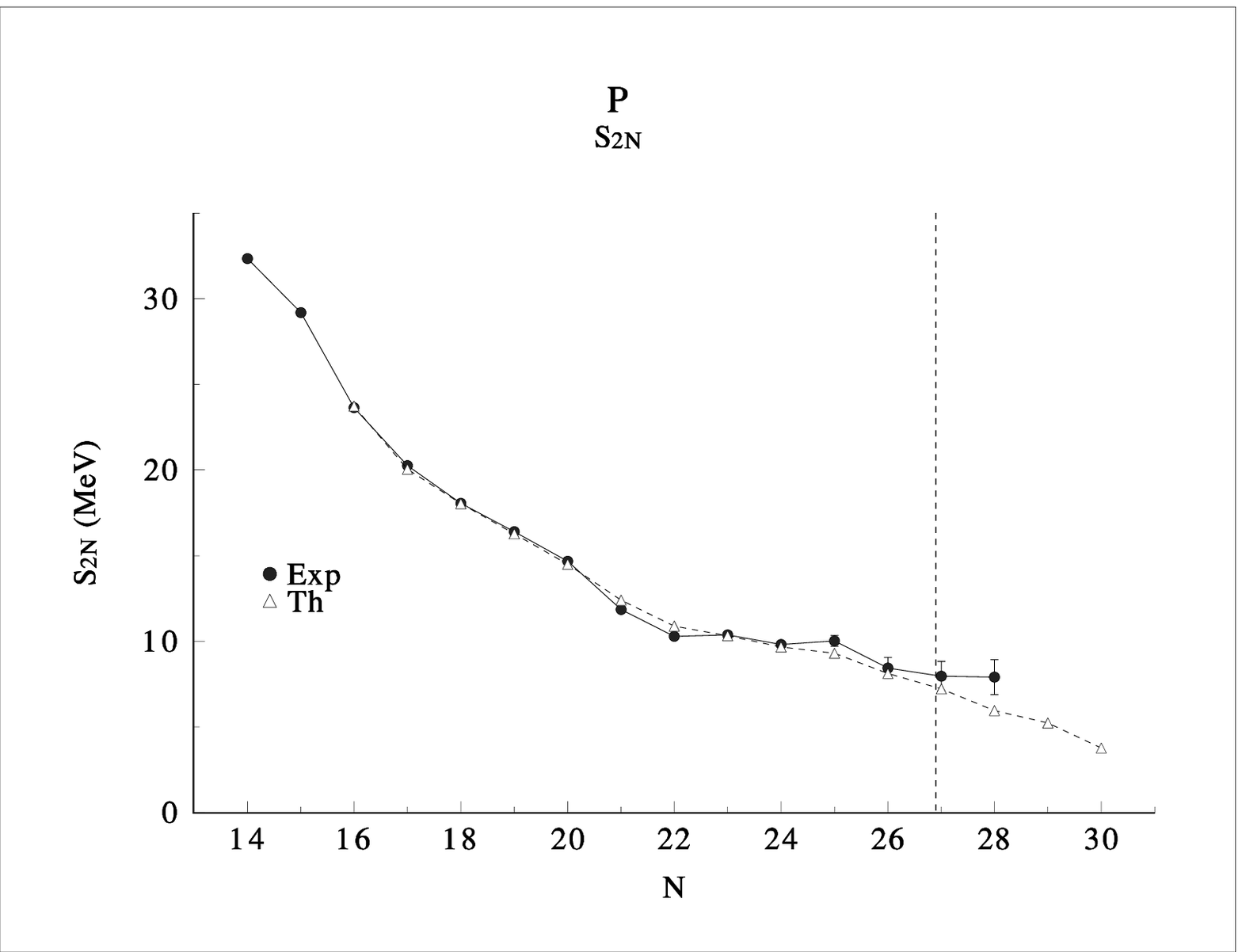,height=9.5cm ,width=8cm}
\end{center}
\end{figure*}

\begin{figure*}[bot]
\begin{center}
\caption[]{S$_{2n}$ energies for the si isotopes. See caption to figure
           \ref{f10}.}
\label{f14}
\vspace*{0.5cm}
\leavevmode
\psfig{file=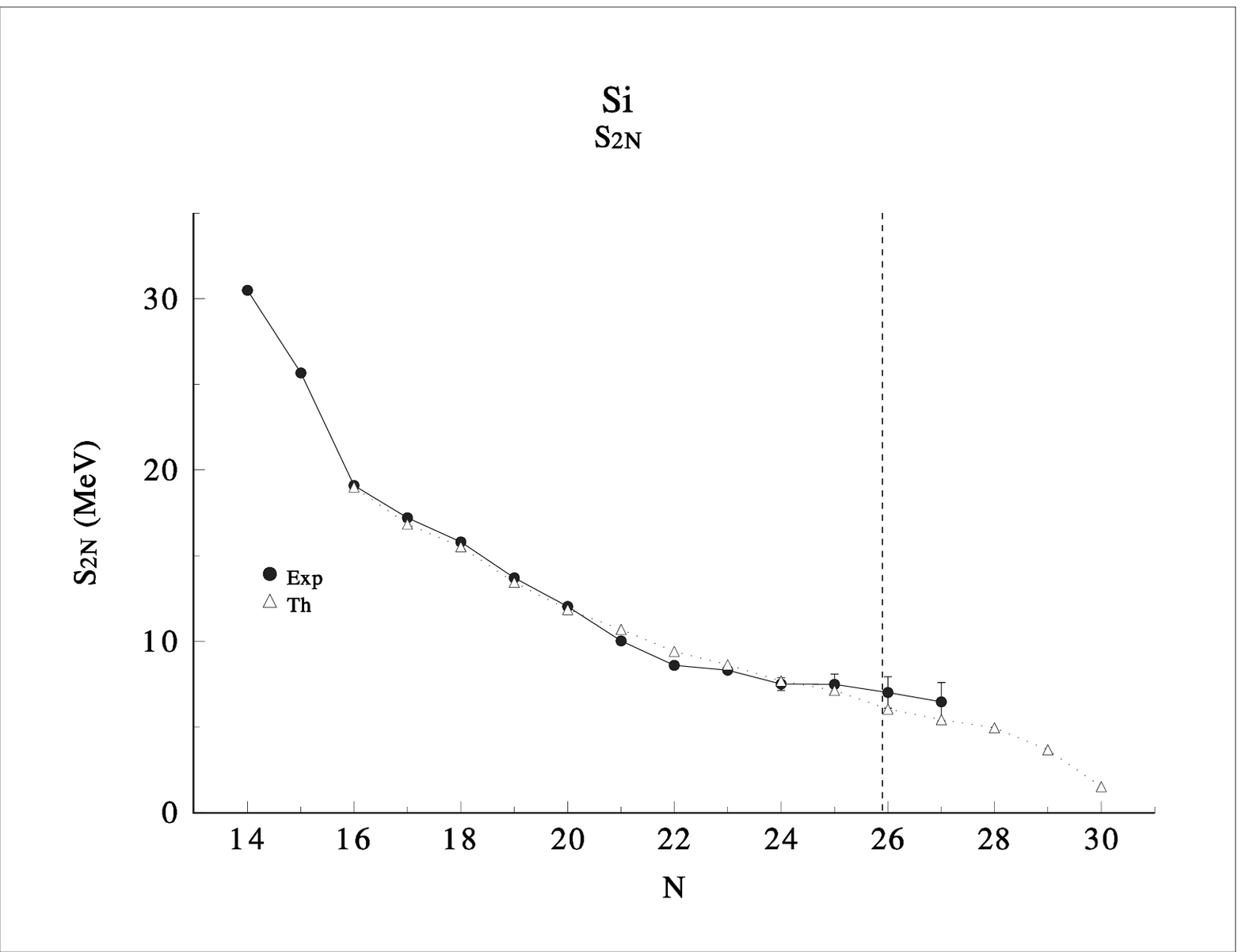,height=9.5cm,width=8cm}
\end{center} 
\end{figure*}

Figure \ref{f13} shows the evolution of the 2$^{+}$ energy for the Ca
and Si isotopes. The large energy increase of the 2$^{+}$ for the four
nuclei $^{40}$Ca, $^{48}$Ca, $^{34}$S and $^{42}$Si is something that
typifies doubly magic nuclei. Notice, however, that the effect is
smaller for the exotic $^{42}$Si.  A similar evidence comes from the
1f$_{7/2}$ occupation numbers along N=28. Those, shown in figure
\ref{f14}, reveal an increase of the f$_{7/2}$ occupation at Z=20 and
Z=14, again slightly in $^{42}$Si.

\begin{figure*}[top]
\begin{center}
\caption[]{2$^{+}$ excitation energies for the Ca and Si isotopes.}
\label{f15}
\vspace*{0.5cm}
\leavevmode
\psfig{file=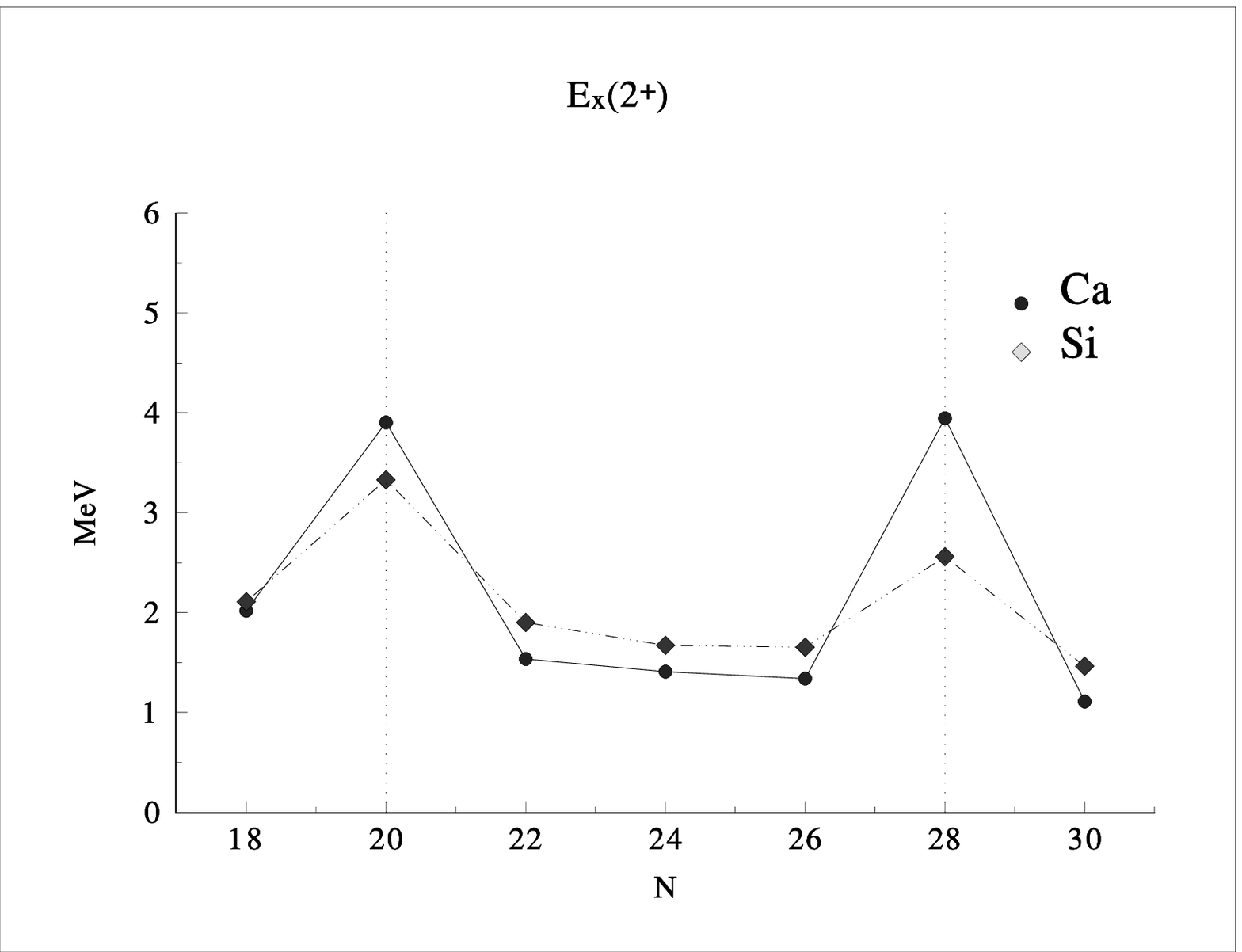,height=9.5cm ,width=8cm}
\end{center}
\end{figure*}

The 2$^{+}$ excitation energies of the isotopes of Ar and S, around
N=28, are presented fig. \ref{f15} and compared with the corresponding
energies of nuclei much closer to the stability at those neutron
numbers (Ti, Cr).  There is an upward shift of the excitation energies
at N=28 that is a little smaller in Ar and S. The behaviour is seen to
be very similar close and far off stability, thus indicating that the
N=28 closure stands in the very neutron rich region.

From the analysis of these three sets of results: S$_{2N}$,
E$_x(2^{+})$ and the occupation numbers we can conclude that the
1f$_{7/2}^{8}$ neutron configuration at N=28, although somewhat eroded
by the large neutron excess, is still dominant and that $^{42}$Si is a
new doubly magic nuclei.  Therefore, the vanishing of the shell
closure far from stability that happens at N=20, does not seem to
occur at N=28.

\subsection{The deformed nuclei $^{40}$S and $^{42}$S}

Our calculations show the existence of low lying 2$^{+}$ states for
the Ar and S isotopes with $22\le$N$\le26$. However, only the Sulphur
isotopes have large (and negative) spectroscopic quadrupole moments
and enhanced E2 transitions (see figs. 18 and 19). $^{42}$Ar and
$^{44}$Ar have reasonably large BE2's but very small spectroscopic
quadrupole moments.

The lowest 2$^{+}$ energies correspond to $^{40}$S and $^{42}$S. Their
spectroscopic quadrupole moments are -17.1e$^2$fm$^2$ and
-19.2e$^2$fm$^2$ respectively.  Assuming that the Bohr-Mottelson
formula for the intrinsic quadrupole moment is applicable, we get
Q$_0$=59.8 e$^2$fm$^2$ for $^{40}$S and Q$_0$=67.2 e$^2$fm$^2$ for
$^{42}$S, corresponding to deformation parameters $\beta \simeq$ 0.29
and $\beta\simeq$ 0.32.  Our results are consistent with the
Relativistic Mean Field calculations of Werner {\em et al.}, which
predict $\beta\simeq 0.25$.  Their non-relativistic calculations point
to a more complicated situation with nearly degenerate prolate
($\beta\simeq 0.25$) and oblate ($\beta\simeq -0.24$) solutions. The
only indication of coexistence in our results is provided by the
presence of a very low (1.51 MeV) second 0$^+$ state in $^{44}$S.

The values of the E2 reduced transition probabilities are 77.9
e$^2$fm$^4$ (A=40) and 93.0 e$^2$fm$^4$ (A=42).  We compare the
quadrupole moments of $^{42}$S and other nuclei near N=28 in figure
\ref{f14}.  The value for $^{42}$S is about 2/3 of that predicted for
$^{48}$Cr leading to very similar deformation parameters $\beta=$0.32
and $\beta=$0.30.  In addition, the yrast levels of $^{42}$S follow
the rotational law J(J+1) up to J=6.

It is also important to notice that, although a truncated 1f$_{7/2}$
2p$_{3/2}$ valence space for the neutrons, contains the basic degrees
of freedom for the description of deformed nuclei in this region, it
enhances either the rotational behaviour or the closed shell features,
depending on the basic configurations that are involved. As an example
let consider the evolution of the $2^{+}$ excitation energies of the
Sulphur isotopes.  In the truncated space the energy of the $2^{+}$ in
N=26 goes down 200 keV to 0.88 MeV while in N=28 it goes up 300 keV to
1.78 MeV.

In order to explain the onset of deformation we shall explore which
are the configurations that are responsible for the quadrupole
coherence in our space.  It is well known that the SU(3) symmetry is
broken by the strong $\vec{l}\cdot\vec{s}$ coupling in the pf shell.
Yet, it has been recently shown \cite{Zuk:95} that the subspaces of a
major shell spanned by $\Delta j=2$ orbits develop an approximate
SU(3) symmetry, named quasi-SU(3). For instance the 1f$_{7/2}$
2p$_{3/2}$ valence space is enough to to explain the collective
behaviour of nuclei in the $pf$ shell.  In this space the maximal
quadrupole collectivity is obtained for configurations with 4 protons
and 4 neutrons.  In our case, the basic neutron degrees of freedom are
contained in this space, but protons are confined in the $sd$ shell.
For the nuclei we are interested in, the orbit 1d$_{5/2}$  is
essentially closed and only the 1d$_{3/2}$ and the 2s$_{1/2}$  can
be considered   active.  In this situation neither SU(3) nor
quasi-SU(3) can develop.  

However, one must look more carefully at the behaviour of these two
orbits.  As far as the neutrons are filling the $sd$ shell, both
orbits remain well separated. For instance, the excitation energy of
the $3/2_{1}$ level in $^{29}$P and $^{35}$P are 1.39 and 2.62 MeV
respectively.  Thus, the ground state wave functions are dominated by
configurations with the maximum allowed number of particles in the
2s$_{1/2}$ orbit.  Particularly, for $^{36}$S the (2s$_{1/2})^4$
configuration represents the 89$\%$ of the wave function.  When the
neutrons begin to fill the $pf$ shell, the two orbitals become more
and more degenerate (see table \ref{t1}) inducing strong mixing in the
wave functions.  A valence space with degenerate 1d$_{3/2}$ and
2s$_{1/2}$ orbitals has the geometry of pseudo-SU(3), and maximal
quadrupole coherence is obtained for two protons, i.e, for the Sulphur
isotopes.  For three particles the quadrupole moment  is strictly
zero and beyond it changes of sign leading to oblate shapes.  Thus, we
expect other nuclei as, for instance, the Ar isotopes to have small
colectivity.  

Notice, however, that strict degeneration is achieved
only at N=28, where the neutron collectivity is rather small.  In
other words, there is a mismatch between  the region where protons have
collective wave functions (N=28) and the region were neutrons have
quadrupole collectivity, that  prevents the existence of a larger region
of deformation. A compromise seems to take place for Z=16 and  N=26 
($^{42}$S) or N=24 ($^{40}$S).

\section{Summary} 
 
In this paper we have studied the N=28 region far from the stability.
We have built an effective interaction for $sd$ active protons and
$pf$ active neutrons that is the natural generalization of successful
$sd$-only and $pf$-only interactions. The agreement of the theoretical
results and the experimental data is very satisfactory.  However, much
more experimental information is needed to obtain a complete
understanding of the region. Accurate information on the $2^{+}$
excitation energies, on the evolution of the $3/2^+,1/2^+$ doublet
along the Cl and P chains, the single particle spectrum of $^{35}$Si
are challenges for the physics of exotic nuclei.  We have also
addressed the question of the magicity of N=28.  A detailed analysis
of the $S_{2n}$, $2^{+}$ energies and occupation numbers allow us to
conclude that it persists in the very neutron rich regime, contrary to
what happens at N=20. Finally our results show that the $^{40,42}$S
isotopes are well deformed, but the effect seems to be very local; no
large region of deformation is therefore expected.

\medskip

{\bf Acknowledgements.} We want to thank A. Zuker his decisive help
in the determination of the effective interaction and G. Walter for
enlightening  discussions.
This work has been partially supported by the
IN2P3(France)-CICyT(Spain) agreements and by DGICyT (Spain) grant
PB93-263.

\clearpage

\begin{table}[top]
\begin{center} 
\caption[]{2$^{+}_1$ excitation energies (in MeV).}
\label{t1}
\vspace*{0.5cm}
\begin{math}\begin{array}{l|cccccc}\hline\hline
\multicolumn{7}{c}{}\\[-6pt]
    &  N       & 22   &   24   &   26    &   28    &   30  \\[2pt]\hline
\multicolumn{7}{c}{}\\[-6pt]
Cr  & th  &0.869 & 0.742  &  0.793  &  1.621   & 0.877 \\[2pt]
    & exp &0.869 & 0.751  &  0.783  &  1.434   & 0.835 \\ [2pt]\hline
Ti  & th  &1.209 & 0.869  &  1.003  &  1.755   & 1.103 \\[2pt]
    & exp &1.083 & 0.889  &  0.983  &  1.555   & 1.047 \\ [2pt]\hline
Ca  & th  &1.534 & 1.409  &  1.338  &  3.947   & 1.110 \\[2pt]
    & exp &1.524 & 1.157  &  1.347  &  3.832   & 1.030 \\[2pt]\hline
Ar  & th  &1.368 & 1.249  &  1.182  &  1.655   & 1.219 \\[2pt]
    & exp &1.461 & 1.208  &    -    &    -     &   -   \\[2pt]\hline
S   & th  &1.209 & 1.047  &  1.073  &  1.465   & 1.272 \\[2pt]
    & exp &  -   &    -   &    -    &     -    &   -   \\[2pt]\hline
Si  & th  &1.900 & 1.670  &  1.652  &  2.558   & 1.461 \\[2pt]
    & exp &  -   &   -    &    -    &     -    &   -   \\[2pt]\hline
\end{array}\end{math}
\end{center}
\end{table}

\begin{table}[here]
\begin{center}
\caption[]{ Energy difference between the  3/2$^{+}_1$ and 1/2$^{+}_1$
 states in K, Cl and P.}
\label{t2}
\vspace*{0.5cm}
\begin{math}\begin{array}{l|lcccccc}\hline\hline
\multicolumn{8}{c}{}\\[-6pt]
    & N  &   20  &  22  &  24    &   26    &   28    &   30  \\[2pt]\hline
\multicolumn{8}{c}{}\\[-6pt] 
K&th  &2.73&0.96 & 0.67  &  0.28 & -0.38 & 0.19 \\[2pt]
E(1/2)-E(3/2) &exp &2.52&0.98 & 0.56  &  0.47 & -0.36 & -     \\[2pt]
Cl&th &1.82&0.71 & 0.06  & -0.21 & -0.05 & -0.06\\[2pt]
E(1/2)-E(3/2) &exp & 1.73 &0.40 &   -   &   -   &  -  &    -  \\[2pt]
P&th &2.61 &1.32 & 0.68  &  0.32 &  0.06 & 0.38     \\[2pt]
E(3/2)-E(1/2) &exp &  -   &  - &   -   &   -   &   -  &    -  \\[2pt]
\end{array}\end{math}
\end{center}
\end{table}

\begin{table}[bot] 
\begin{center}
\caption[]{Total Gamow-Teller strength and theoretical vs.  measured
 half-lives of the even-even Ar and  S isotopes.}
\label{t3}
\vspace*{0.5cm}
\begin{math}\begin{array}{cccc}\hline\hline
   &       &                      &      \\
   &GT^{-} &T^{th}_{\frac{1}{2}}  & T^{exp}_{\frac{1}{2}}
   \\[5pt]\hline 
   &       &                      &      \\
^{44}Ar&22.85&4.70 m&11.87 m\\[2pt] 
^{46}Ar&28.84&6.94 s&8.4 s\\[2pt] 
^{38}S&16.78&174.4 m&170.3 m\\ [2pt]
^{40}S&22.87&6.67 s&8.8 s\\[2pt] 
^{42}S&28.89&389 ms&560 ms\\[2pt] 
^{44}S&34.93&54 ms&123 ms\\[2pt]\hline 
\end{array}\end{math} 
\end{center} 
\end{table} 

\clearpage

\clearpage

\clearpage

\clearpage

\clearpage

\clearpage

\clearpage

\clearpage

\begin{figure*}[bot]
\begin{center}
\caption[]{2$^{+}$ excitation energies for the Ca and Si isotopes.}
\label{f16}
\vspace*{0.5cm}
\leavevmode
\psfig{file=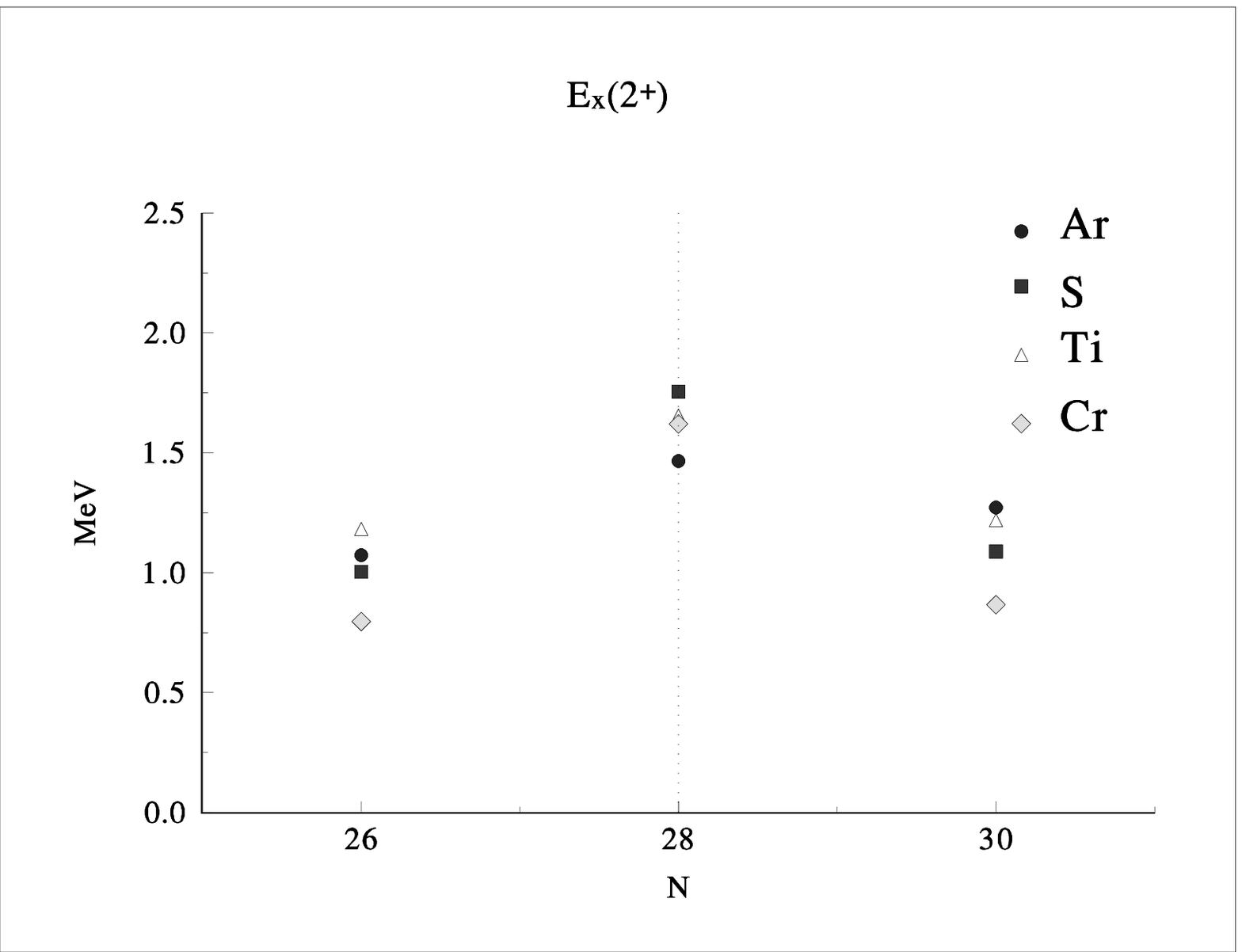,height=9.5cm,width=8cm}
\end{center}
\end{figure*}

\clearpage

\begin{figure*}[top]
\begin{center}
\caption[]{f$_{7/2}$ occupation numbers.}
\label{f17}
\vspace*{0.5cm}
\leavevmode
\psfig{file=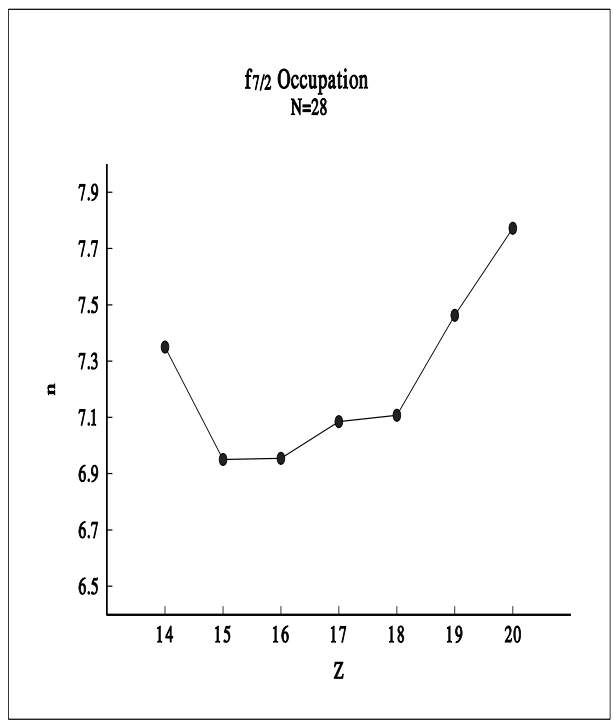,height=9.5cm ,width=8cm}
\end{center}
\end{figure*}

\begin{figure*}[bot]
\begin{center}
\caption[]{Spectroscopic Quadrupole moments for the Cr,Ti,Ar and S isotopes.}
\label{f18}
\vspace*{0.5cm}
\leavevmode
\psfig{file=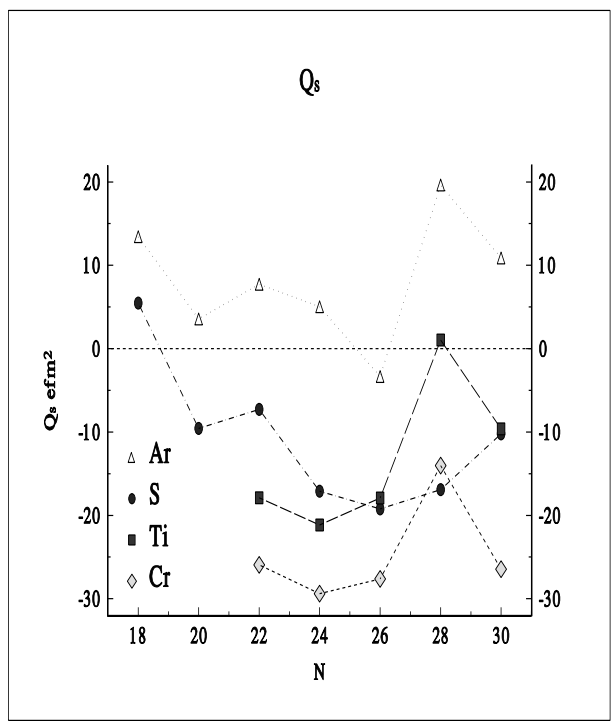,height=9.5cm,width=8cm}
\end{center}
\end{figure*}

\begin{figure*}[bot]
\begin{center}
\caption[]{BE2 for  Ca, Ar, S and Si isotopes.}
\label{f19}
\vspace*{0.5cm}
\leavevmode
\psfig{file=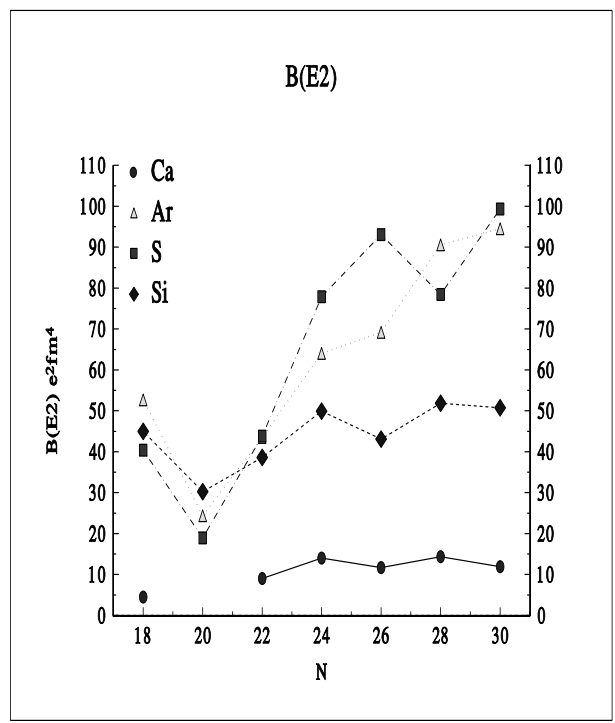,height=9.5cm,width=8cm}
\end{center}
\end{figure*}

\end{document}